\documentclass[prl,twocolumn,superscriptaddress,showkeys,showpacs]{revtex4-1}

\usepackage{graphicx}% Include figure files
\usepackage{bm}% bold math
\usepackage{amsmath}
\usepackage{amssymb}
\begin{document}

\title{The influence of the structural transition on magnetic fluctuations in NaFeAs}

\author{Juanjuan Liu}
\affiliation{Department of Physics, Renmin University of China, Beijing 100872, China}

\author{Jinchen Wang}
\affiliation{Department of Physics, Renmin University of China, Beijing 100872, China}

\author{Wei Luo}
\affiliation{Department of Physics, Renmin University of China, Beijing 100872, China}

\author{Jieming Sheng}
\affiliation{Department of Physics, Renmin University of China, Beijing 100872, China}

\author{Aifeng Wang}
\affiliation{Hefei National Laboratory for Physical Sciences at Microscale and Department of Physics, University of Science and Technology of China, Hefei, Anhui
230026, China}

\author{Xianhui Chen}
\affiliation{Hefei National Laboratory for Physical Sciences at Microscale and Department of Physics, University of Science and Technology of China, Hefei, Anhui
230026, China}
\affiliation{Collaborative Innovation Center of Advanced Microstructures, Nanjing University, Nanjing 210093, China}

\author{Sergey A. Danilkin}
\affiliation{Bragg Institute, ANSTO, Lucas Heights, NSW 2234, Australia}

\author{Wei Bao}
\email{wbao@ruc.edu.cn}
\affiliation{Department of Physics, Renmin University of China, Beijing 100872, China}

\begin{abstract}
NaFeAs belongs to a class of Fe-based superconductors which parent compounds show separated structural and magnetic transitions. Effects of the structural transition on spin dynamics therefore can be investigated separately from the magnetic transition. A plateau in dynamic spin response is observed in a critical region around the structural transition temperature $T_S$. It is interpreted as due to the stiffening of spin fluctuations along the in-plane magnetic hard axis due to the $d_{xz}$ and $d_{yz}$ orbital ordering. The appearance of anisotropic spin dynamics in the critical region above the $T_S$ at $T^*$ offers a dynamic magnetic scattering mechanism for anisotropic electronic properties in the commonly referred ``nematic phase''.
\end{abstract}

%74.70.Xa	Pnictides and chalcogenides
%75.25.Dk 	Orbital, charge, and other orders,including coupling of these orders
%75.40.Gb	Dynamic properties (dynamic susceptibility, spin waves, spin diffusion, dynamic scaling, etc.)
%78.70.Nx 	Neutron inelastic scattering
\maketitle

%%%%%%%%%%%%%%%%%%%%%%%%%%%%%%%%%%%%%%%%%%%%%%%%%%%%%%%%%%

It is well known that in the phase diagrams of the LaFeAsO (1111) \cite{Kamihara2008}, BaFe$_2$As$_2$ (122) \cite{A054630} and FeSe (11) \cite{A072369} superconductors, there exists an antiferromagnetic order in a distorted lattice phase which breaks the four-fold tetragonal symmetry \cite{A040795,A062776,A092058}.
For all parent antiferromagnetic orders of the 1111, 122 and 11 families \cite{A062195,A062776,A092058} as well as the antiferromagnetic order of K$_2$Fe$_4$Se$_5$ \cite{D020830}, neutron diffraction experiments, which simultaneously measure crystalline and magnetic structures, have shown that the antiferromagnetic bond between neighboring Fe ions has expanded lattice spacing while the ferromagnetic bond has contracted lattice spacing. Such a close relation between the sign of magnetic interaction and the lattice spacing is a hallmark of the orbital ordering phenomenon \cite{A062195,A062776,A092058,D020830,F087405,rev_mit}. Together with theoretical calculation \cite{A042252}, it has been further stated specifically in 2008 that the orbital ordering involves the $d_{xz}$ and $d_{yz}$ orbitals in the Fe-based material \cite{A062195}. The two types of Fe bonds of different $d_{xz}$ and $d_{yz}$ occupancy have since been shown to account for the three kinds of commensurate antiferromagnetic structures which have been observed so far for the Fe-based superconductor parent compounds \cite{F087405,E060881}, and different $d_{xz}$ and $d_{yz}$ occupancy in the structural distorted phase has also been subsequently observed in ARPES experiments \cite{Yi11_pnas}.

When the number of electrons is less than available $d$ orbitals, an ordered occupation pattern of the $d$ orbitals by the electrons leads to a regular orbital-pair pattern \cite{F087405,E060881}. The hopping parameter between the transition-metal ion pair determines not only the magnetic exchange interaction, but also the lattice bonding strength, thus the lattice spacing, as well as the transport property. This was demonstrated by Goodenough \cite{orb_ge} in his classic explanation of the rich magnetic phases in perovskite manganites observed in neutron diffraction experiments \cite{wollan}. Orbital ordering has also been identified through such lattice and magnetic interaction corresponding relation more recently in neutron scattering studies on classic transition metal oxides \cite{bao96c,bao96b}. Since the orbital ordering changes the bonding between the transition-metal ions, it automatically manifests itself by a structural transition \cite{bao96c,bao96b}.
The effective spin Hamiltonian is also altered by the orbital ordering transition. When (I) the
N\'{e}el temperature $T_N$ of the effective spin Hamiltonian in the orbital ordered state is higher than the orbital order temperature $T_S$, the antiferromagnetic transition will necessarily concur with the structural transition, such as in the case of V$_2$O$_3$ \cite{bao96c}, BaFe$_2$As$_2$ \cite{A062776} and FeTe \cite{A092058}.
When (II) $T_N$ is lower than $T_S$, the antiferromagnetic transition will occur in a separated phase transition upon further cooling after the structural transition, such as in the case of the $C$-type perovskite manganite \cite{bao96b}, LaFeAsO \cite{A040795} and NaFeAs (111) \cite{Chen09_prl,Wang12_prb,Li09_prb}.

Since the antiferromagnetic bond along the $a$-axis and the ferromagnetic bond along the $b$-axis
have different spacing \cite{A062776}, detwinned 122 crystals would reveal the different hopping parameters of the two different bonds in transport measurements along the $a$ and $b$-axis \cite{Chu10_science,Tanatar10_prb}. The in-plane anisotropy expected to occur in the orbital ordered state has also been observed in resonant ultrasound spectroscopy \cite{Fernandes10_prl}, torque magnetometry \cite{Kasahara12_nat}, magnetic inelastic neutron scattering \cite{Lu14_science}, time-resolved polarimetry \cite{Patz14_natcomm} measurements of the 122 systems. However, theoretical debate has been going on concerning the relative importance of the orbital \cite{E060881,Lv09_prb,Lee09_prl,Chen10_prb,Lee12_prb,ZouLJ_prb}, lattice \cite{Liang13_prl} and spin \cite{Fang08_prb,Xu08_prb} degrees of freedom, and a spontaneous symmetry breaking Fermi liquid state, the so-called ``nematic phase'', has been introduced \cite{Fernandes14_NatPhys}. In this connection, the class II orbital-ordered Fe-based superconductors such as the LaFeAsO \cite{A040795} and NaFeAs \cite{Li09_prb} families of separated $T_N$ and $T_S$ offer experimental advantage. LaFeAsO and Co-doped BaFe$_2$As$_2$ have been investigated in NMR and inelastic neutron scattering studies to reveal the effect of the structural transition on spin dynamics \cite{Fu12_prl,Zhang15_prl}. Here we report inelastic neutron scattering investigation on NaFeAs.

\begin{figure}[bt!]
	\includegraphics[width=\columnwidth,keepaspectratio]{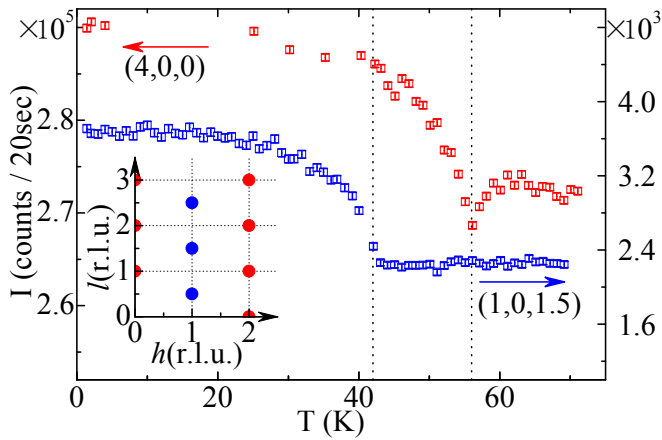}
	\vskip -.2cm
	\caption{(color online). The peak intensity of the nuclear Bragg peak (4,0,0) (red) and magnetic Bragg peak (1,0,1.5) (blue) as a function of temperature. 
		The two vertical dash lines denote the structure and magnetic transitions, respectively. 
		Inset: The $(h0l)$ reciprocal plane with the red circles marking the nuclear Bragg spots, 
		and blue the magnetic Bragg spots.}
	\label{ord} 
\end{figure}

Single crystals were synthesized as described in \cite{Wang12_prb}. 
About 2.5 grams of samples were co-aligned with the mosaic $\sim$$3^{\circ}$ 
in the $(h0l)$ scattering plane of the low-temperature orthorhombic unit cell. 
Both nuclear and magnetic Bragg peaks are accessible in this plane \cite{Li09_prb}, see inset to Fig.~\ref{ord}.
The lattice parameters $a = 5.590$, $b=5.570$ and $c = 6.993 \mathring{A}$ at 1.5 K.
Neutron scattering experiments were performed at the thermal neutron triple-axis spectrometer Taipan \cite{taipan} in Bragg Institute, Australia Nuclear Science and Technology Organization (ANSTO).
The Pyrolytic Graphite (PG) monochromator was in the double focusing mode and the PG analyzer in the vertical focusing mode.
The final energy of the neutron beam was fixed at 14.7 meV, a PG filter was used after the sample to remove higher order neutrons and a 40$^{\prime}$ collimator was put after the sample.
The sample temperature was regulated using an ILL Orange cryostat in the 1.4 K to 300 K range. 
Fig.~\ref{ord} shows the temperature dependence of the nuclear Bragg peak (4,0,0) and magnetic Bragg peak (1,0,1.5). 
The anomaly of the nuclear peak marks the the tetragonal to orthorhombic structural transition at $T_S =56$ K, and the magnetic peak appears below the N\'{e}el temperature $T_{N} = 42$ K. The two phase transitions are well separated.

%%%%%%%%%%%%%%%%%%%%%%%%%%%%%%%%%%%%%%%%%%%%%%%%%%%%%%%%%%

\begin{figure}[tb!]
\includegraphics[width=\columnwidth]{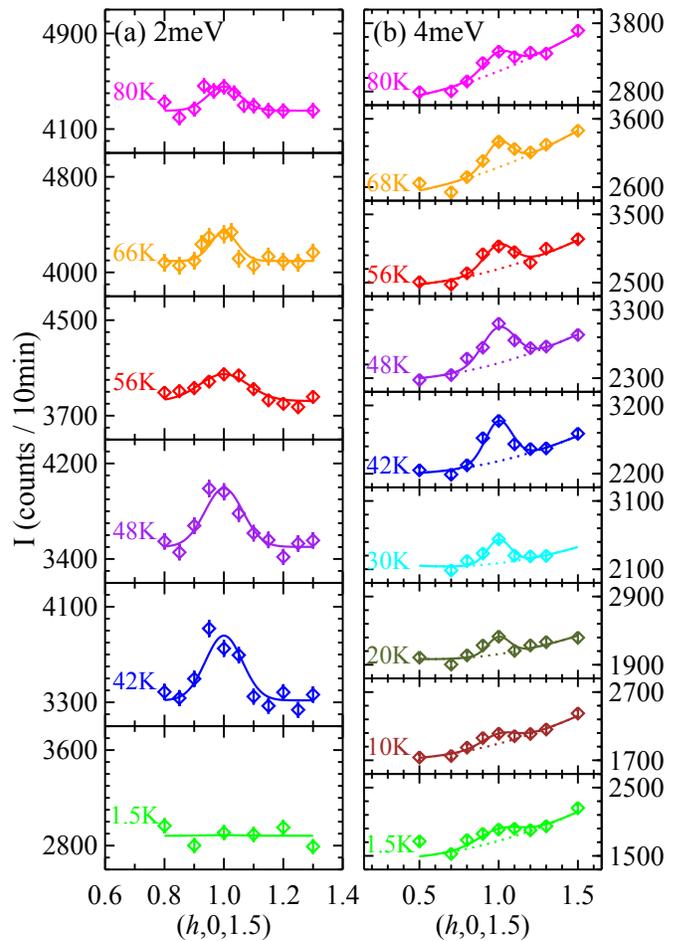}
\vskip -.2cm
\caption{(color online). 
	Constant energy scan across the magnetic zone center (1,0,1.5) at (a) 2 meV  and (b) 4 meV at various temperature from 1.4 K to 80 K. 
}
\label{hscan} 
\end{figure}

The constant energy scans along the $a$-axis across the magnetic zone center (1,0,1.5) are displayed in Fig.~\ref{hscan}. 
At $E=2$ meV, no peak was visible at 1.5 K, consistent with a spin gap formation due to the long-range magnetic order \cite{Song13_prb}. 
The most intense peak appears at $T_N= 42$ K when the gap closes and magnetic critical scattering maximizes.
The low energy magnetic fluctuation persists into the paramagnetic state, both below and above the structural transition at $T_S=56$ K. 
At $E=4$ meV, a finite peak at the magnetic zone center can be detected at the base temperature 1.4 K.
Its intensity increases upon approaching the $T_N$, then decreases upon further rising of temperature, as expected for magnetic excitations.

\begin{figure}[tb!]
\includegraphics[width=0.75\columnwidth]{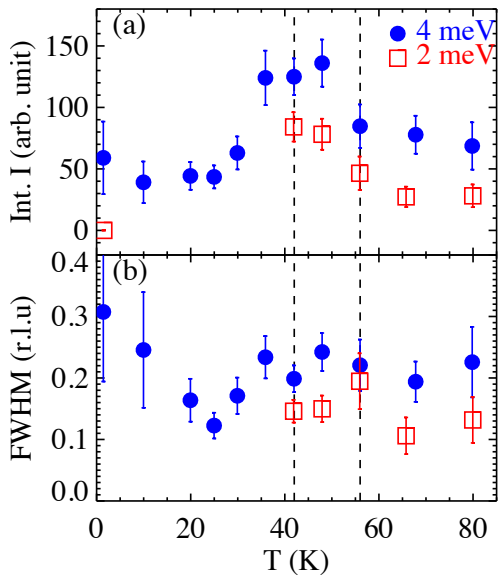}
\vskip -.2cm
\caption{(color online). 
(a) The integrated intensity and (b) the full-width-at-half-maximum (FWHM) in the reciprocal lattice unit of the constant energy scan in Fig.~\ref{hscan} at 2 meV (red square) and 4 meV (blue circle). The two vertical dash lines denote the structural and magnetic transitions of NaFeAs, respectively. 
}
\label{fig3}
\end{figure}

The fitting parameters of these constant-$E$ scans across (1,0,1.5) are shown in Fig.~\ref{fig3}.
A comparison with the results of similar scans in a recent neutron scattering study on LaFeAsO and Ba(Fe$_{0.953}$Co$_{0.047}$)$_2$As$_2$ \cite{Zhang15_prl} would be beneficial.
The maximum of the magnetic signal at $T_N$ in Fig.~\ref{fig3}(a) is the same as the cases for LaFeAsO and Ba(Fe$_{0.953}$Co$_{0.047}$)$_2$As$_2$. However, this is a universal critical phenomenon of the second order antiferromagnetic transition, which occurs in any class II orbital ordered material. Such a critical behavior has also been picked up in the NMR study on NaFeAs \cite{Ma11_prb}.
For LaFeAsO and Ba(Fe$_{0.953}$Co$_{0.047}$)$_2$As$_2$, the peak width reduces drastically below $T_S$ by $\sim 0.12(4) \AA^{-1}$ and $0.07(2) \AA^{-1}$, respectively \cite{Zhang15_prl}.
For NaFeAs, however, 
our data statistics does not allow such a solid conclusion, putting the upper limit of the peak narrowing at $\sim$0.04 $\AA^{-1}$.

\begin{figure}[tb!]
\includegraphics[width=0.75\columnwidth]{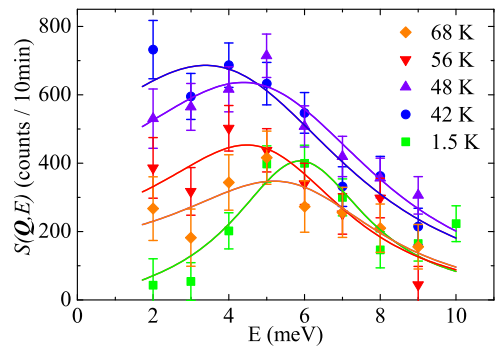}
\vskip -.2cm
\caption{(color online). The dynamic magnetic correlation function $S({\bf Q}, E)$ at the magnetic Bragg wave vector 
${\bf Q}=(1,0,1.5)$ as a function of energy at various temperatures. The background has been subtracted, as described in text.
The energy gap at 1.5 K closes when temperature is raised above $T_N$.
}
\label{fig4} 
\end{figure}

Due to the hydrogen-containing glue used in assembling the single crystals, it contributed substantial background scattering, such as those shown in Fig.~\ref{hscan} beneath the magnetic peak. Thus, in order to extract magnetic signal $S({\bf Q}, E)$ from the inelastic neutron scattering signal $I$, we performed const-$E$ scans from 1.4 to 80 K and covered the energy range up to 10 meV. Magnetic signal in our investigation range is sharp enough in the reciprocal ${\bf Q}$ space, it reaches background level at $h=0.7$ and 1.3 in the const-$E$ scan along the ($h$,0,1.5). Therefore we obtain $S({\bf Q}, E)$ at ${\bf Q}=(1,0,1.5)$ from measurement of $I({\bf Q}, E)$ at ${\bf Q}=(1,0,1.5)$ subtracted by background measured at ${\bf Q}=(0.7,0,1.5)$ and ${\bf Q}=(1.3,0,1.5)$.
Fig.~\ref{fig4} shows $S({\bf Q}, E)$ at the peak position ${\bf Q}=(1,0,1.5)$ in energy scans at various temperatures. There is an energy gap at 1.5 K, consistent with the measurement from a previous neutron scattering experiment at 2 K \cite{Song13_prb}. At $T_N =42$ K, the gap is closed and an overdamped critical spin dynamic response appears. The critical spin dynamics extend above the structural or orbital transition at $T_S$.

\begin{figure}[tb!]
\includegraphics[width=\columnwidth]{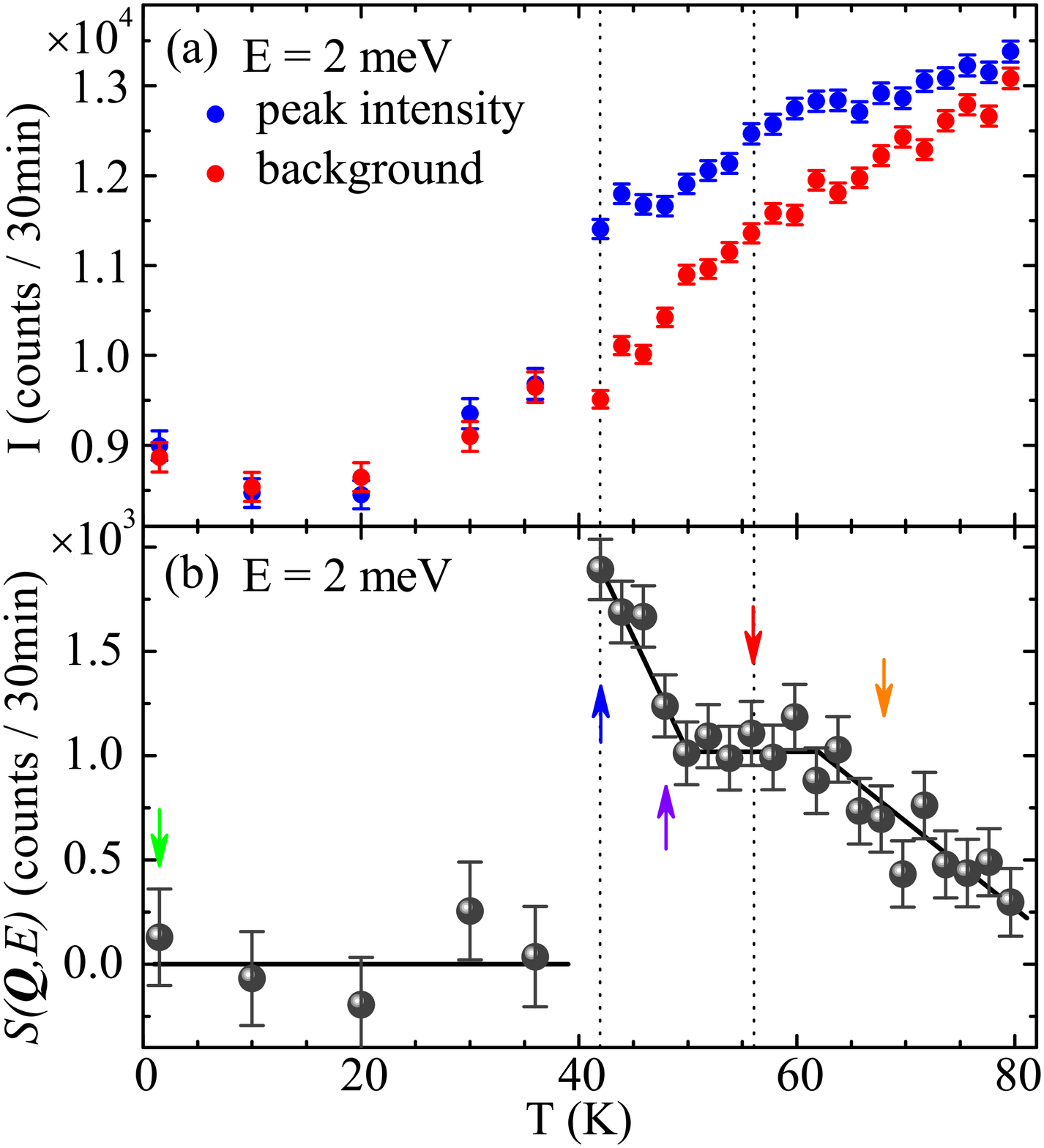}
\vskip -.2cm
\caption{(color online). (a) The peak intensity (blue circle) of the const.-$E=2$ meV scan at the magnetic zone center ${\bf Q}=(1,0,1.5)$ and background (red circle) at ${\bf Q}=(1.3,0,1.5)$ as a function of temperature. 
(b) The dynamic magnetic correlation function $S({\bf Q}, E)$ at ${\bf Q}=(1,0,1.5)$ and $E=2$ meV as a function of temperature. 
The vertical dash lines mark the magnetic and structural transition temperatures at $T_{N} = 42$ K and $T_S =56$ K, respectively. The color arrows in (b) indicate the temperatures at which the const-$E$ scans in Fig.~2 and const-${\bf Q}$ scans in Fig.~4 are shown with symbols in the same colors. The plateau of $S({\bf Q}, E)$ around $T_S$ is likely duo to the suppression of the spin fluctuations along the magnetic hard axis along the $b$-axis. The upper end of the plateau is $T^*$, below which anisotropy in spin dynamics and consequently anisotropic scatterings of conduction electrons occur.
}
\label{fig5} 
\end{figure}

To pin down the elusive influence of the structural transition on spin dynamics, we now focus on the most temperature sensitive part of dynamic magnetic correlation function $S({\bf Q}, E)$ at ${\bf Q}=(1,0,1.5)$ and $E=2$ meV. Fig.~\ref{fig5}(a) shows the raw data of the peak intensity and background measured at ${\bf Q}=(1.3,0,1.5)$ as a function of temperature. The longer counting time was used to ensure adequate data statistics for the background subtracted $S({\bf Q}, E)$ shown in Fig.~\ref{fig5}(b). The energy gap at 1.5 K, demonstrated in Fig.~\ref{hscan}(a) and Fig.~\ref{fig4}, disappears rather abruptly upon warming the sample to the N\'{e}el temperature $T_N$, where low energy spin fluctuations peak and then decrease upon further warming the sample.
The otherwise $\lambda$-shaped signal of magnetic critical fluctuations is modified prominently by a plateau around the $T_S$. The abnormal behavior is rather puzzling, in particularly if one follows the building up of the magnetic correlations from high temperature to the base temperature: more and more spins join the dynamic magnetic correlations at the magnetic wave vector, as signified by the increasing intensity with lowering the temperature. However, the build-up is arrested in the neighborhood of the tetragonal-to-orthorhombic structural transition $|T-T_N| \le 6$ K. Only with further lowering the temperature, the dynamic spin correlations resume the normal magnetic behavior with increasing intensity, which finally condense to the long-range antiferromagnetic order at $T_N$ and gap out the low energy spin fluctuations.

While our measurements shown in Fig.~\ref{fig5} were performed at a low energy in the energy gap, a possible explanation of the arrest of the spin dynamics at the plateau may be provided from the polarized neutron scattering results at 6 meV above the gap \cite{Song13_prb}. Song et al.\ show that while spin fluctuations above $T_S$ are isotropic, the transverse component along the ferromagnetic bond direction is partially frozen out at $T_S$. If such a behavior holds also below the gap energy caused by the spin-space anisotropy, one may explain the reduced intensity in the plateau region of our data. Thus, the orbital ordering renders the ferromagnetic bond along the $b$-axis the magnetic hard axis through the usual spin-orbital coupling mechanism, consistent with the observed alignment of the magnetic moments along the easy axis in the $a$-axis discovered in the 1111 \cite{A062195}, 122 \cite{A062776} and 111 \cite{Li09_prb} families of Fe-based materials.

Resistivity anisotropy along the $a$ and $b$-axis has also been observed in the NaFeAs family of the Fe-based superconductors \cite{Deng15_prb} similar to the 122 family materials \cite{Chu10_science,Tanatar10_prb}, as expected for orbital ordered materials. Such a behavior usually starts above the $T_S$ at a higher temperature $T^*$.
Anisotropic electronic state has also been deduced from the 
quasiparticle interference (QPI) measurements \cite{Rosenthal14_natphys,Cai14_prl} and observed in ARPES measurements \cite{He10_prl,Zhang12_prb,Yi12_njp} on the 111 materials. The $T^*$ is reported to be $\sim$70 K for NaFeAs. If we take the partial frozen picture discussed above, the upper end of the plateau in Fig.~\ref{fig5} indicates the appearance of the spin space anisotropy, namely the $T^* \approx 62$ K. Our $T^*$ is lower than the value $\sim$70 K for two possible reasons: 1) One would expect higher value of $T^*$ if the measurement energy of spin fluctuations is reduced from 2 meV to $0^+$. 2) the uniaxial stress to detwin the sample is similar to the uniform magnetic field applying to a ferromagnet, and it is well known that the field increases the $T_C$. The plateau extends above and below the $T_S$ approximately symmetrically, it is likely attributed to the critical fluctuations of the orbital degree of freedom of the orbital ordering transition at $T_S$.

In summary, we performed inelastic neutron scattering investigation on NaFeAs which shows separated structural and antiferromagnetic transitions upon cooling. 
The prominent influence of the structural transition on spin dynamics manifests in the plateau in the temperature dependence of the low energy dynamic magnetic fluctuations. It reflects the stiffening of spin fluctuations along the ferromagnetic bond in the $b$-axis and is consistent with the eventual condensation of the dynamic magnetic correlations to the long range antiferromagnetic order with the easy axis along the antiferromagnetic bond direction in the $a$-axis. High statistics data allow us to detect the plateau feature, which could be a consequence of the orbital fluctuations and extends the anisotropic spin fluctuations to $T^*$ above the $T_S$. Therefore, an orbital ordering picture involving the $d_{xz}$ and $d_{yz}$ orbitals which we have proposed since 2008 \cite{A062195,A062776} basing on the pioneering idea of Goodenough \cite{orb_ge} seems to consistently explain all experimental data on NaFeAs.

%\section{Conclusion}

%\section{Acknowledgements}
The works at RUC and USTC were supported by National Basic Research Program of China 
(Grant Nos. 2012CB921700 and 2011CBA00112) and the National Natural Science
Foundation of China (Grant Nos. 11034012 and 11190024).

%\bibliography{NaFeAs,/home/wei/kept/tex/bib4/FeAs,/home/wei/kept/tex/bib4/mine}

\begin{thebibliography}{46}%
\makeatletter
\providecommand \@ifxundefined [1]{%
 \@ifx{#1\undefined}
}%
\providecommand \@ifnum [1]{%
 \ifnum #1\expandafter \@firstoftwo
 \else \expandafter \@secondoftwo
 \fi
}%
\providecommand \@ifx [1]{%
 \ifx #1\expandafter \@firstoftwo
 \else \expandafter \@secondoftwo
 \fi
}%
\providecommand \natexlab [1]{#1}%
\providecommand \enquote  [1]{``#1''}%
\providecommand \bibnamefont  [1]{#1}%
\providecommand \bibfnamefont [1]{#1}%
\providecommand \citenamefont [1]{#1}%
\providecommand \href@noop [0]{\@secondoftwo}%
\providecommand \href [0]{\begingroup \@sanitize@url \@href}%
\providecommand \@href[1]{\@@startlink{#1}\@@href}%
\providecommand \@@href[1]{\endgroup#1\@@endlink}%
\providecommand \@sanitize@url [0]{\catcode `\\12\catcode `\$12\catcode
  `\&12\catcode `\#12\catcode `\^12\catcode `\_12\catcode `\%12\relax}%
\providecommand \@@startlink[1]{}%
\providecommand \@@endlink[0]{}%
\providecommand \url  [0]{\begingroup\@sanitize@url \@url }%
\providecommand \@url [1]{\endgroup\@href {#1}{\urlprefix }}%
\providecommand \urlprefix  [0]{URL }%
\providecommand \Eprint [0]{\href }%
\providecommand \doibase [0]{http://dx.doi.org/}%
\providecommand \selectlanguage [0]{\@gobble}%
\providecommand \bibinfo  [0]{\@secondoftwo}%
\providecommand \bibfield  [0]{\@secondoftwo}%
\providecommand \translation [1]{[#1]}%
\providecommand \BibitemOpen [0]{}%
\providecommand \bibitemStop [0]{}%
\providecommand \bibitemNoStop [0]{.\EOS\space}%
\providecommand \EOS [0]{\spacefactor3000\relax}%
\providecommand \BibitemShut  [1]{\csname bibitem#1\endcsname}%
\let\auto@bib@innerbib\@empty
%</preamble>
\bibitem [{\citenamefont {Kamihara}\ \emph {et~al.}(2008)\citenamefont
  {Kamihara}, \citenamefont {Watanabe}, \citenamefont {Hirano},\ and\
  \citenamefont {Hosono}}]{Kamihara2008}%
  \BibitemOpen
  \bibfield  {author} {\bibinfo {author} {\bibfnamefont {Y.}~\bibnamefont
  {Kamihara}}, \bibinfo {author} {\bibfnamefont {T.}~\bibnamefont {Watanabe}},
  \bibinfo {author} {\bibfnamefont {M.}~\bibnamefont {Hirano}}, \ and\ \bibinfo
  {author} {\bibfnamefont {H.}~\bibnamefont {Hosono}},\ }\href@noop {}
  {\bibfield  {journal} {\bibinfo  {journal} {J.\ Am.\ Chem.\ Soc.}\ }\textbf
  {\bibinfo {volume} {130}},\ \bibinfo {pages} {3296} (\bibinfo {year}
  {2008})}\BibitemShut {NoStop}%
\bibitem [{\citenamefont {Rotter}\ \emph {et~al.}(2008)\citenamefont {Rotter},
  \citenamefont {Tegel},\ and\ \citenamefont {Johrendt}}]{A054630}%
  \BibitemOpen
  \bibfield  {author} {\bibinfo {author} {\bibfnamefont {M.}~\bibnamefont
  {Rotter}}, \bibinfo {author} {\bibfnamefont {M.}~\bibnamefont {Tegel}}, \
  and\ \bibinfo {author} {\bibfnamefont {D.}~\bibnamefont {Johrendt}},\
  }\href@noop {} {\bibfield  {journal} {\bibinfo  {journal} {Phys. Rev. Lett.}\
  }\textbf {\bibinfo {volume} {101}},\ \bibinfo {pages} {107006} (\bibinfo
  {year} {2008})}\BibitemShut {NoStop}%
\bibitem [{\citenamefont {Hsu}\ \emph {et~al.}(2008)\citenamefont {Hsu},
  \citenamefont {Luo}, \citenamefont {Yeh}, \citenamefont {Chen}, \citenamefont
  {Huang}, \citenamefont {Wu}, \citenamefont {Lee}, \citenamefont {Huang},
  \citenamefont {Chu}, \citenamefont {Yan},\ and\ \citenamefont
  {Wu}}]{A072369}%
  \BibitemOpen
  \bibfield  {author} {\bibinfo {author} {\bibfnamefont {F.-C.}\ \bibnamefont
  {Hsu}}, \bibinfo {author} {\bibfnamefont {J.-Y.}\ \bibnamefont {Luo}},
  \bibinfo {author} {\bibfnamefont {K.-W.}\ \bibnamefont {Yeh}}, \bibinfo
  {author} {\bibfnamefont {T.-K.}\ \bibnamefont {Chen}}, \bibinfo {author}
  {\bibfnamefont {T.-W.}\ \bibnamefont {Huang}}, \bibinfo {author}
  {\bibfnamefont {P.~M.}\ \bibnamefont {Wu}}, \bibinfo {author} {\bibfnamefont
  {Y.-C.}\ \bibnamefont {Lee}}, \bibinfo {author} {\bibfnamefont {Y.-L.}\
  \bibnamefont {Huang}}, \bibinfo {author} {\bibfnamefont {Y.-Y.}\ \bibnamefont
  {Chu}}, \bibinfo {author} {\bibfnamefont {D.-C.}\ \bibnamefont {Yan}}, \ and\
  \bibinfo {author} {\bibfnamefont {M.-K.}\ \bibnamefont {Wu}},\ }\href@noop {}
  {\bibfield  {journal} {\bibinfo  {journal} {PNAS}\ }\textbf {\bibinfo
  {volume} {105}},\ \bibinfo {pages} {14262} (\bibinfo {year}
  {2008})}\BibitemShut {NoStop}%
\bibitem [{\citenamefont {de~la Cruz}\ \emph {et~al.}(2008)\citenamefont {de~la
  Cruz}, \citenamefont {Huang}, \citenamefont {Lynn}, \citenamefont {Li},
  \citenamefont {Ratcliff}, \citenamefont {Zarestky}, \citenamefont {Mook},
  \citenamefont {Chen}, \citenamefont {Luo}, \citenamefont {Wang},\ and\
  \citenamefont {Dai}}]{A040795}%
  \BibitemOpen
  \bibfield  {author} {\bibinfo {author} {\bibfnamefont {C.}~\bibnamefont
  {de~la Cruz}}, \bibinfo {author} {\bibfnamefont {Q.}~\bibnamefont {Huang}},
  \bibinfo {author} {\bibfnamefont {J.~W.}\ \bibnamefont {Lynn}}, \bibinfo
  {author} {\bibfnamefont {J.}~\bibnamefont {Li}}, \bibinfo {author}
  {\bibfnamefont {W.}~\bibnamefont {Ratcliff}}, \bibinfo {author}
  {\bibfnamefont {J.~L.}\ \bibnamefont {Zarestky}}, \bibinfo {author}
  {\bibfnamefont {H.~A.}\ \bibnamefont {Mook}}, \bibinfo {author}
  {\bibfnamefont {G.~F.}\ \bibnamefont {Chen}}, \bibinfo {author}
  {\bibfnamefont {J.~L.}\ \bibnamefont {Luo}}, \bibinfo {author} {\bibfnamefont
  {N.~L.}\ \bibnamefont {Wang}}, \ and\ \bibinfo {author} {\bibfnamefont
  {P.}~\bibnamefont {Dai}},\ }\href@noop {} {\bibfield  {journal} {\bibinfo
  {journal} {Nature}\ }\textbf {\bibinfo {volume} {453}},\ \bibinfo {pages}
  {899} (\bibinfo {year} {2008})}\BibitemShut {NoStop}%
\bibitem [{\citenamefont {Huang}\ \emph {et~al.}(2008)\citenamefont {Huang},
  \citenamefont {Qiu}, \citenamefont {Bao}, \citenamefont {Green},
  \citenamefont {Lynn}, \citenamefont {Gasparovic}, \citenamefont {Wu},
  \citenamefont {Wu},\ and\ \citenamefont {Chen}}]{A062776}%
  \BibitemOpen
  \bibfield  {author} {\bibinfo {author} {\bibfnamefont {Q.}~\bibnamefont
  {Huang}}, \bibinfo {author} {\bibfnamefont {Y.}~\bibnamefont {Qiu}}, \bibinfo
  {author} {\bibfnamefont {W.}~\bibnamefont {Bao}}, \bibinfo {author}
  {\bibfnamefont {M.}~\bibnamefont {Green}}, \bibinfo {author} {\bibfnamefont
  {J.}~\bibnamefont {Lynn}}, \bibinfo {author} {\bibfnamefont {Y.}~\bibnamefont
  {Gasparovic}}, \bibinfo {author} {\bibfnamefont {T.}~\bibnamefont {Wu}},
  \bibinfo {author} {\bibfnamefont {G.}~\bibnamefont {Wu}}, \ and\ \bibinfo
  {author} {\bibfnamefont {X.~H.}\ \bibnamefont {Chen}},\ }\href@noop {}
  {\bibfield  {journal} {\bibinfo  {journal} {Phys. Rev. Lett.}\ }\textbf
  {\bibinfo {volume} {101}},\ \bibinfo {pages} {257003} (\bibinfo {year}
  {2008})}\BibitemShut {NoStop}%
\bibitem [{\citenamefont {Bao}\ \emph {et~al.}(2009)\citenamefont {Bao},
  \citenamefont {Qiu}, \citenamefont {Huang}, \citenamefont {Green},
  \citenamefont {Zajdel}, \citenamefont {Fitzsimmons}, \citenamefont
  {Zhernenkov}, \citenamefont {Fang}, \citenamefont {Qian}, \citenamefont
  {Vehstedt}, \citenamefont {Yang}, \citenamefont {Pham}, \citenamefont
  {Spinu},\ and\ \citenamefont {Mao}}]{A092058}%
  \BibitemOpen
  \bibfield  {author} {\bibinfo {author} {\bibfnamefont {W.}~\bibnamefont
  {Bao}}, \bibinfo {author} {\bibfnamefont {Y.}~\bibnamefont {Qiu}}, \bibinfo
  {author} {\bibfnamefont {Q.}~\bibnamefont {Huang}}, \bibinfo {author}
  {\bibfnamefont {M.~A.}\ \bibnamefont {Green}}, \bibinfo {author}
  {\bibfnamefont {P.}~\bibnamefont {Zajdel}}, \bibinfo {author} {\bibfnamefont
  {M.~R.}\ \bibnamefont {Fitzsimmons}}, \bibinfo {author} {\bibfnamefont
  {M.}~\bibnamefont {Zhernenkov}}, \bibinfo {author} {\bibfnamefont
  {M.}~\bibnamefont {Fang}}, \bibinfo {author} {\bibfnamefont {B.}~\bibnamefont
  {Qian}}, \bibinfo {author} {\bibfnamefont {E.}~\bibnamefont {Vehstedt}},
  \bibinfo {author} {\bibfnamefont {J.}~\bibnamefont {Yang}}, \bibinfo {author}
  {\bibfnamefont {H.}~\bibnamefont {Pham}}, \bibinfo {author} {\bibfnamefont
  {L.}~\bibnamefont {Spinu}}, \ and\ \bibinfo {author} {\bibfnamefont
  {Z.}~\bibnamefont {Mao}},\ }\href@noop {} {\bibfield  {journal} {\bibinfo
  {journal} {Phys. Rev. Lett.}\ }\textbf {\bibinfo {volume} {102}},\ \bibinfo
  {pages} {247001} (\bibinfo {year} {2009})}\BibitemShut {NoStop}%
\bibitem [{\citenamefont {Qiu}\ \emph {et~al.}(2008)\citenamefont {Qiu},
  \citenamefont {Bao}, \citenamefont {Huang}, \citenamefont {Yildirim},
  \citenamefont {Simmons}, \citenamefont {Green}, \citenamefont {Lynn},
  \citenamefont {Gasparovic}, \citenamefont {Li}, \citenamefont {Wu},
  \citenamefont {Wu},\ and\ \citenamefont {Chen}}]{A062195}%
  \BibitemOpen
  \bibfield  {author} {\bibinfo {author} {\bibfnamefont {Y.}~\bibnamefont
  {Qiu}}, \bibinfo {author} {\bibfnamefont {W.}~\bibnamefont {Bao}}, \bibinfo
  {author} {\bibfnamefont {Q.}~\bibnamefont {Huang}}, \bibinfo {author}
  {\bibfnamefont {T.}~\bibnamefont {Yildirim}}, \bibinfo {author}
  {\bibfnamefont {J.}~\bibnamefont {Simmons}}, \bibinfo {author} {\bibfnamefont
  {M.}~\bibnamefont {Green}}, \bibinfo {author} {\bibfnamefont
  {J.}~\bibnamefont {Lynn}}, \bibinfo {author} {\bibfnamefont {Y.}~\bibnamefont
  {Gasparovic}}, \bibinfo {author} {\bibfnamefont {J.}~\bibnamefont {Li}},
  \bibinfo {author} {\bibfnamefont {T.}~\bibnamefont {Wu}}, \bibinfo {author}
  {\bibfnamefont {G.}~\bibnamefont {Wu}}, \ and\ \bibinfo {author}
  {\bibfnamefont {X.}~\bibnamefont {Chen}},\ }\href@noop {} {\bibfield
  {journal} {\bibinfo  {journal} {Phys. Rev. Lett.}\ }\textbf {\bibinfo
  {volume} {101}},\ \bibinfo {pages} {257002} (\bibinfo {year}
  {2008})}\BibitemShut {NoStop}%
\bibitem [{\citenamefont {Bao}\ \emph {et~al.}(2011)\citenamefont {Bao},
  \citenamefont {Huang}, \citenamefont {Chen}, \citenamefont {Green},
  \citenamefont {Wang}, \citenamefont {He}, \citenamefont {Wang},\ and\
  \citenamefont {Qiu}}]{D020830}%
  \BibitemOpen
  \bibfield  {author} {\bibinfo {author} {\bibfnamefont {W.}~\bibnamefont
  {Bao}}, \bibinfo {author} {\bibfnamefont {Q.}~\bibnamefont {Huang}}, \bibinfo
  {author} {\bibfnamefont {G.~F.}\ \bibnamefont {Chen}}, \bibinfo {author}
  {\bibfnamefont {M.~A.}\ \bibnamefont {Green}}, \bibinfo {author}
  {\bibfnamefont {D.~M.}\ \bibnamefont {Wang}}, \bibinfo {author}
  {\bibfnamefont {J.~B.}\ \bibnamefont {He}}, \bibinfo {author} {\bibfnamefont
  {X.~Q.}\ \bibnamefont {Wang}}, \ and\ \bibinfo {author} {\bibfnamefont
  {Y.}~\bibnamefont {Qiu}},\ }\href@noop {} {\bibfield  {journal} {\bibinfo
  {journal} {Chin. Phys. Lett.}\ }\textbf {\bibinfo {volume} {28}},\ \bibinfo
  {pages} {086104} (\bibinfo {year} {2011})}\BibitemShut {NoStop}%
\bibitem [{\citenamefont {Bao}(2013)}]{F087405}%
  \BibitemOpen
  \bibfield  {author} {\bibinfo {author} {\bibfnamefont {W.}~\bibnamefont
  {Bao}},\ }\href@noop {} {\bibfield  {journal} {\bibinfo  {journal} {Chinese
  Phys. B}\ }\textbf {\bibinfo {volume} {22}},\ \bibinfo {pages} {087405}
  (\bibinfo {year} {2013})}\BibitemShut {NoStop}%
\bibitem [{\citenamefont {Imada}\ \emph {et~al.}(1998)\citenamefont {Imada},
  \citenamefont {Fujimori},\ and\ \citenamefont {Tokura}}]{rev_mit}%
  \BibitemOpen
  \bibfield  {author} {\bibinfo {author} {\bibfnamefont {M.}~\bibnamefont
  {Imada}}, \bibinfo {author} {\bibfnamefont {A.}~\bibnamefont {Fujimori}}, \
  and\ \bibinfo {author} {\bibfnamefont {Y.}~\bibnamefont {Tokura}},\
  }\href@noop {} {\bibfield  {journal} {\bibinfo  {journal} {Rev. Mod. Phys.}\
  }\textbf {\bibinfo {volume} {70}},\ \bibinfo {pages} {1039} (\bibinfo {year}
  {1998})}\BibitemShut {NoStop}%
\bibitem [{\citenamefont {Yildirim}(2008)}]{A042252}%
  \BibitemOpen
  \bibfield  {author} {\bibinfo {author} {\bibfnamefont {T.}~\bibnamefont
  {Yildirim}},\ }\href@noop {} {\bibfield  {journal} {\bibinfo  {journal}
  {Phys. Rev. Lett.}\ }\textbf {\bibinfo {volume} {101}},\ \bibinfo {pages}
  {057010} (\bibinfo {year} {2008})}\BibitemShut {NoStop}%
\bibitem [{\citenamefont {Yin}\ \emph {et~al.}(2012)\citenamefont {Yin},
  \citenamefont {Lin},\ and\ \citenamefont {Ku}}]{E060881}%
  \BibitemOpen
  \bibfield  {author} {\bibinfo {author} {\bibfnamefont {W.-G.}\ \bibnamefont
  {Yin}}, \bibinfo {author} {\bibfnamefont {C.-H.}\ \bibnamefont {Lin}}, \ and\
  \bibinfo {author} {\bibfnamefont {W.}~\bibnamefont {Ku}},\ }\href@noop {}
  {\bibfield  {journal} {\bibinfo  {journal} {Phys. Rev. B}\ }\textbf {\bibinfo
  {volume} {86}},\ \bibinfo {pages} {081106(R)} (\bibinfo {year}
  {2012})}\BibitemShut {NoStop}%
\bibitem [{\citenamefont {Yi}\ \emph {et~al.}(2011)\citenamefont {Yi},
  \citenamefont {Lu}, \citenamefont {Chu}, \citenamefont {Analytis},
  \citenamefont {Sorini}, \citenamefont {Kemper}, \citenamefont {Moritz},
  \citenamefont {Mo}, \citenamefont {Moore}, \citenamefont {Hashimoto},
  \citenamefont {Lee}, \citenamefont {Hussain}, \citenamefont {Devereaux},
  \citenamefont {Fisher},\ and\ \citenamefont {Shen}}]{Yi11_pnas}%
  \BibitemOpen
  \bibfield  {author} {\bibinfo {author} {\bibfnamefont {M.}~\bibnamefont
  {Yi}}, \bibinfo {author} {\bibfnamefont {D.}~\bibnamefont {Lu}}, \bibinfo
  {author} {\bibfnamefont {J.-H.}\ \bibnamefont {Chu}}, \bibinfo {author}
  {\bibfnamefont {J.~G.}\ \bibnamefont {Analytis}}, \bibinfo {author}
  {\bibfnamefont {A.~P.}\ \bibnamefont {Sorini}}, \bibinfo {author}
  {\bibfnamefont {A.~F.}\ \bibnamefont {Kemper}}, \bibinfo {author}
  {\bibfnamefont {B.}~\bibnamefont {Moritz}}, \bibinfo {author} {\bibfnamefont
  {S.-K.}\ \bibnamefont {Mo}}, \bibinfo {author} {\bibfnamefont {R.~G.}\
  \bibnamefont {Moore}}, \bibinfo {author} {\bibfnamefont {M.}~\bibnamefont
  {Hashimoto}}, \bibinfo {author} {\bibfnamefont {W.-S.}\ \bibnamefont {Lee}},
  \bibinfo {author} {\bibfnamefont {Z.}~\bibnamefont {Hussain}}, \bibinfo
  {author} {\bibfnamefont {T.~P.}\ \bibnamefont {Devereaux}}, \bibinfo {author}
  {\bibfnamefont {I.~R.}\ \bibnamefont {Fisher}}, \ and\ \bibinfo {author}
  {\bibfnamefont {Z.-X.}\ \bibnamefont {Shen}},\ }\href@noop {} {\bibfield
  {journal} {\bibinfo  {journal} {PNAS}\ }\textbf {\bibinfo {volume} {108}},\
  \bibinfo {pages} {6878} (\bibinfo {year} {2011})}\BibitemShut {NoStop}%
\bibitem [{\citenamefont {Goodenough}(1955)}]{orb_ge}%
  \BibitemOpen
  \bibfield  {author} {\bibinfo {author} {\bibfnamefont {J.~B.}\ \bibnamefont
  {Goodenough}},\ }\href@noop {} {\bibfield  {journal} {\bibinfo  {journal}
  {Phys. Rev.}\ }\textbf {\bibinfo {volume} {100}},\ \bibinfo {pages} {564}
  (\bibinfo {year} {1955})}\BibitemShut {NoStop}%
\bibitem [{\citenamefont {Wollan}\ and\ \citenamefont
  {Koehler}(1955)}]{wollan}%
  \BibitemOpen
  \bibfield  {author} {\bibinfo {author} {\bibfnamefont {E.~O.}\ \bibnamefont
  {Wollan}}\ and\ \bibinfo {author} {\bibfnamefont {W.~C.}\ \bibnamefont
  {Koehler}},\ }\href@noop {} {\bibfield  {journal} {\bibinfo  {journal} {Phys.
  Rev.}\ }\textbf {\bibinfo {volume} {100}},\ \bibinfo {pages} {545} (\bibinfo
  {year} {1955})}\BibitemShut {NoStop}%
\bibitem [{\citenamefont {Bao}\ \emph {et~al.}(1997{\natexlab{a}})\citenamefont
  {Bao}, \citenamefont {Broholm}, \citenamefont {Aeppli}, \citenamefont {Dai},
  \citenamefont {Honig},\ and\ \citenamefont {Metcalf}}]{bao96c}%
  \BibitemOpen
  \bibfield  {author} {\bibinfo {author} {\bibfnamefont {W.}~\bibnamefont
  {Bao}}, \bibinfo {author} {\bibfnamefont {C.}~\bibnamefont {Broholm}},
  \bibinfo {author} {\bibfnamefont {G.}~\bibnamefont {Aeppli}}, \bibinfo
  {author} {\bibfnamefont {P.}~\bibnamefont {Dai}}, \bibinfo {author}
  {\bibfnamefont {J.~M.}\ \bibnamefont {Honig}}, \ and\ \bibinfo {author}
  {\bibfnamefont {P.}~\bibnamefont {Metcalf}},\ }\href@noop {} {\bibfield
  {journal} {\bibinfo  {journal} {Phys. Rev. Lett.}\ }\textbf {\bibinfo
  {volume} {78}},\ \bibinfo {pages} {507} (\bibinfo {year}
  {1997}{\natexlab{a}})}\BibitemShut {NoStop}%
\bibitem [{\citenamefont {Bao}\ \emph {et~al.}(1997{\natexlab{b}})\citenamefont
  {Bao}, \citenamefont {Axe}, \citenamefont {Chen},\ and\ \citenamefont
  {Cheong}}]{bao96b}%
  \BibitemOpen
  \bibfield  {author} {\bibinfo {author} {\bibfnamefont {W.}~\bibnamefont
  {Bao}}, \bibinfo {author} {\bibfnamefont {J.~D.}\ \bibnamefont {Axe}},
  \bibinfo {author} {\bibfnamefont {C.~H.}\ \bibnamefont {Chen}}, \ and\
  \bibinfo {author} {\bibfnamefont {S.-W.}\ \bibnamefont {Cheong}},\
  }\href@noop {} {\bibfield  {journal} {\bibinfo  {journal} {Phys. Rev. Lett.}\
  }\textbf {\bibinfo {volume} {78}},\ \bibinfo {pages} {543} (\bibinfo {year}
  {1997}{\natexlab{b}})}\BibitemShut {NoStop}%
\bibitem [{\citenamefont {Chen}\ \emph {et~al.}(2009)\citenamefont {Chen},
  \citenamefont {Hu}, \citenamefont {Luo},\ and\ \citenamefont
  {Wang}}]{Chen09_prl}%
  \BibitemOpen
  \bibfield  {author} {\bibinfo {author} {\bibfnamefont {G.~F.}\ \bibnamefont
  {Chen}}, \bibinfo {author} {\bibfnamefont {W.~Z.}\ \bibnamefont {Hu}},
  \bibinfo {author} {\bibfnamefont {J.~L.}\ \bibnamefont {Luo}}, \ and\
  \bibinfo {author} {\bibfnamefont {N.~L.}\ \bibnamefont {Wang}},\ }\href
  {\doibase 10.1103/PhysRevLett.102.227004} {\bibfield  {journal} {\bibinfo
  {journal} {Phys. Rev. Lett.}\ }\textbf {\bibinfo {volume} {102}},\ \bibinfo
  {pages} {227004} (\bibinfo {year} {2009})}\BibitemShut {NoStop}%
\bibitem [{\citenamefont {Wang}\ \emph {et~al.}(2012)\citenamefont {Wang},
  \citenamefont {Luo}, \citenamefont {Yan}, \citenamefont {Ying}, \citenamefont
  {Xiang}, \citenamefont {Ye}, \citenamefont {Cheng}, \citenamefont {Li},
  \citenamefont {Hu},\ and\ \citenamefont {Chen}}]{Wang12_prb}%
  \BibitemOpen
  \bibfield  {author} {\bibinfo {author} {\bibfnamefont {A.~F.}\ \bibnamefont
  {Wang}}, \bibinfo {author} {\bibfnamefont {X.~G.}\ \bibnamefont {Luo}},
  \bibinfo {author} {\bibfnamefont {Y.~J.}\ \bibnamefont {Yan}}, \bibinfo
  {author} {\bibfnamefont {J.~J.}\ \bibnamefont {Ying}}, \bibinfo {author}
  {\bibfnamefont {Z.~J.}\ \bibnamefont {Xiang}}, \bibinfo {author}
  {\bibfnamefont {G.~J.}\ \bibnamefont {Ye}}, \bibinfo {author} {\bibfnamefont
  {P.}~\bibnamefont {Cheng}}, \bibinfo {author} {\bibfnamefont {Z.~Y.}\
  \bibnamefont {Li}}, \bibinfo {author} {\bibfnamefont {W.~J.}\ \bibnamefont
  {Hu}}, \ and\ \bibinfo {author} {\bibfnamefont {X.~H.}\ \bibnamefont
  {Chen}},\ }\href {\doibase 10.1103/PhysRevB.85.224521} {\bibfield  {journal}
  {\bibinfo  {journal} {Phys. Rev. B}\ }\textbf {\bibinfo {volume} {85}},\
  \bibinfo {pages} {224521} (\bibinfo {year} {2012})}\BibitemShut {NoStop}%
\bibitem [{\citenamefont {Li}\ \emph {et~al.}(2009)\citenamefont {Li},
  \citenamefont {de~la Cruz}, \citenamefont {Huang}, \citenamefont {Chen},
  \citenamefont {Xia}, \citenamefont {Luo}, \citenamefont {Wang},\ and\
  \citenamefont {Dai}}]{Li09_prb}%
  \BibitemOpen
  \bibfield  {author} {\bibinfo {author} {\bibfnamefont {S.}~\bibnamefont
  {Li}}, \bibinfo {author} {\bibfnamefont {C.}~\bibnamefont {de~la Cruz}},
  \bibinfo {author} {\bibfnamefont {Q.}~\bibnamefont {Huang}}, \bibinfo
  {author} {\bibfnamefont {G.~F.}\ \bibnamefont {Chen}}, \bibinfo {author}
  {\bibfnamefont {T.-L.}\ \bibnamefont {Xia}}, \bibinfo {author} {\bibfnamefont
  {J.~L.}\ \bibnamefont {Luo}}, \bibinfo {author} {\bibfnamefont {N.~L.}\
  \bibnamefont {Wang}}, \ and\ \bibinfo {author} {\bibfnamefont
  {P.}~\bibnamefont {Dai}},\ }\href {\doibase 10.1103/PhysRevB.80.020504}
  {\bibfield  {journal} {\bibinfo  {journal} {Phys. Rev. B}\ }\textbf {\bibinfo
  {volume} {80}},\ \bibinfo {pages} {020504} (\bibinfo {year}
  {2009})}\BibitemShut {NoStop}%
\bibitem [{\citenamefont {Chu}\ \emph {et~al.}(2010)\citenamefont {Chu},
  \citenamefont {Analytis}, \citenamefont {Greve}, \citenamefont {McMahon},
  \citenamefont {Islam}, \citenamefont {Yamamoto},\ and\ \citenamefont
  {Fisher}}]{Chu10_science}%
  \BibitemOpen
  \bibfield  {author} {\bibinfo {author} {\bibfnamefont {J.-H.}\ \bibnamefont
  {Chu}}, \bibinfo {author} {\bibfnamefont {J.~G.}\ \bibnamefont {Analytis}},
  \bibinfo {author} {\bibfnamefont {K.~D.}\ \bibnamefont {Greve}}, \bibinfo
  {author} {\bibfnamefont {P.~L.}\ \bibnamefont {McMahon}}, \bibinfo {author}
  {\bibfnamefont {Z.}~\bibnamefont {Islam}}, \bibinfo {author} {\bibfnamefont
  {Y.}~\bibnamefont {Yamamoto}}, \ and\ \bibinfo {author} {\bibfnamefont
  {I.~R.}\ \bibnamefont {Fisher}},\ }\href@noop {} {\bibfield  {journal}
  {\bibinfo  {journal} {Science}\ }\textbf {\bibinfo {volume} {329}},\ \bibinfo
  {pages} {824} (\bibinfo {year} {2010})}\BibitemShut {NoStop}%
\bibitem [{\citenamefont {Tanatar}\ \emph {et~al.}(2010)\citenamefont
  {Tanatar}, \citenamefont {Blomberg}, \citenamefont {Kreyssig}, \citenamefont
  {Kim}, \citenamefont {Ni}, \citenamefont {Thaler}, \citenamefont {Bud’ko},
  \citenamefont {Canfield}, \citenamefont {Goldman}, \citenamefont {Mazin},\
  and\ \citenamefont {Prozorov}}]{Tanatar10_prb}%
  \BibitemOpen
  \bibfield  {author} {\bibinfo {author} {\bibfnamefont {M.~A.}\ \bibnamefont
  {Tanatar}}, \bibinfo {author} {\bibfnamefont {E.~C.}\ \bibnamefont
  {Blomberg}}, \bibinfo {author} {\bibfnamefont {A.}~\bibnamefont {Kreyssig}},
  \bibinfo {author} {\bibfnamefont {M.~G.}\ \bibnamefont {Kim}}, \bibinfo
  {author} {\bibfnamefont {N.}~\bibnamefont {Ni}}, \bibinfo {author}
  {\bibfnamefont {A.}~\bibnamefont {Thaler}}, \bibinfo {author} {\bibfnamefont
  {S.~L.}\ \bibnamefont {Bud’ko}}, \bibinfo {author} {\bibfnamefont {P.~C.}\
  \bibnamefont {Canfield}}, \bibinfo {author} {\bibfnamefont {A.~I.}\
  \bibnamefont {Goldman}}, \bibinfo {author} {\bibfnamefont {I.~I.}\
  \bibnamefont {Mazin}}, \ and\ \bibinfo {author} {\bibfnamefont
  {R.}~\bibnamefont {Prozorov}},\ }\href@noop {} {\bibfield  {journal}
  {\bibinfo  {journal} {Phys. Rev. B}\ }\textbf {\bibinfo {volume} {81}},\
  \bibinfo {pages} {184508} (\bibinfo {year} {2010})}\BibitemShut {NoStop}%
\bibitem [{\citenamefont {Fernandes}\ \emph {et~al.}(2010)\citenamefont
  {Fernandes}, \citenamefont {VanBebber}, \citenamefont {Bhattacharya},
  \citenamefont {Chandra}, \citenamefont {Keppens}, \citenamefont {Mandrus},
  \citenamefont {McGuire}, \citenamefont {Sales}, \citenamefont {Sefat},\ and\
  \citenamefont {Schmalian}}]{Fernandes10_prl}%
  \BibitemOpen
  \bibfield  {author} {\bibinfo {author} {\bibfnamefont {R.~M.}\ \bibnamefont
  {Fernandes}}, \bibinfo {author} {\bibfnamefont {L.~H.}\ \bibnamefont
  {VanBebber}}, \bibinfo {author} {\bibfnamefont {S.}~\bibnamefont
  {Bhattacharya}}, \bibinfo {author} {\bibfnamefont {P.}~\bibnamefont
  {Chandra}}, \bibinfo {author} {\bibfnamefont {V.}~\bibnamefont {Keppens}},
  \bibinfo {author} {\bibfnamefont {D.}~\bibnamefont {Mandrus}}, \bibinfo
  {author} {\bibfnamefont {M.~A.}\ \bibnamefont {McGuire}}, \bibinfo {author}
  {\bibfnamefont {B.~C.}\ \bibnamefont {Sales}}, \bibinfo {author}
  {\bibfnamefont {A.~S.}\ \bibnamefont {Sefat}}, \ and\ \bibinfo {author}
  {\bibfnamefont {J.}~\bibnamefont {Schmalian}},\ }\href@noop {} {\bibfield
  {journal} {\bibinfo  {journal} {Phys. Rev. Lett.}\ }\textbf {\bibinfo
  {volume} {105}},\ \bibinfo {pages} {157003} (\bibinfo {year}
  {2010})}\BibitemShut {NoStop}%
\bibitem [{\citenamefont {Kasahara}\ \emph {et~al.}(2012)\citenamefont
  {Kasahara}, \citenamefont {Shi}, \citenamefont {Hashimoto}, \citenamefont
  {Tonegawa}, \citenamefont {Mizukami}, \citenamefont {Shibauchi},
  \citenamefont {Sugimoto}, \citenamefont {Fukuda}, \citenamefont {Terashima},
  \citenamefont {Nevidomskyy},\ and\ \citenamefont {Matsuda}}]{Kasahara12_nat}%
  \BibitemOpen
  \bibfield  {author} {\bibinfo {author} {\bibfnamefont {S.}~\bibnamefont
  {Kasahara}}, \bibinfo {author} {\bibfnamefont {H.~J.}\ \bibnamefont {Shi}},
  \bibinfo {author} {\bibfnamefont {K.}~\bibnamefont {Hashimoto}}, \bibinfo
  {author} {\bibfnamefont {S.}~\bibnamefont {Tonegawa}}, \bibinfo {author}
  {\bibfnamefont {Y.}~\bibnamefont {Mizukami}}, \bibinfo {author}
  {\bibfnamefont {T.}~\bibnamefont {Shibauchi}}, \bibinfo {author}
  {\bibfnamefont {K.}~\bibnamefont {Sugimoto}}, \bibinfo {author}
  {\bibfnamefont {T.}~\bibnamefont {Fukuda}}, \bibinfo {author} {\bibfnamefont
  {T.}~\bibnamefont {Terashima}}, \bibinfo {author} {\bibfnamefont {A.~H.}\
  \bibnamefont {Nevidomskyy}}, \ and\ \bibinfo {author} {\bibfnamefont
  {Y.}~\bibnamefont {Matsuda}},\ }\href@noop {} {\bibfield  {journal} {\bibinfo
   {journal} {Nature}\ }\textbf {\bibinfo {volume} {486}},\ \bibinfo {pages}
  {382} (\bibinfo {year} {2012})}\BibitemShut {NoStop}%
\bibitem [{\citenamefont {Lu}\ \emph {et~al.}(2014)\citenamefont {Lu},
  \citenamefont {Park}, \citenamefont {Zhang}, \citenamefont {Luo},
  \citenamefont {Nevidomskyy}, \citenamefont {Si},\ and\ \citenamefont
  {Dai}}]{Lu14_science}%
  \BibitemOpen
  \bibfield  {author} {\bibinfo {author} {\bibfnamefont {X.}~\bibnamefont
  {Lu}}, \bibinfo {author} {\bibfnamefont {J.~T.}\ \bibnamefont {Park}},
  \bibinfo {author} {\bibfnamefont {R.}~\bibnamefont {Zhang}}, \bibinfo
  {author} {\bibfnamefont {H.}~\bibnamefont {Luo}}, \bibinfo {author}
  {\bibfnamefont {A.~H.}\ \bibnamefont {Nevidomskyy}}, \bibinfo {author}
  {\bibfnamefont {Q.}~\bibnamefont {Si}}, \ and\ \bibinfo {author}
  {\bibfnamefont {P.}~\bibnamefont {Dai}},\ }\href@noop {} {\bibfield
  {journal} {\bibinfo  {journal} {Science}\ }\textbf {\bibinfo {volume}
  {345}},\ \bibinfo {pages} {657} (\bibinfo {year} {2014})}\BibitemShut
  {NoStop}%
\bibitem [{\citenamefont {Patz}\ \emph {et~al.}(2014)\citenamefont {Patz},
  \citenamefont {Li}, \citenamefont {Ran}, \citenamefont {Fernandes},
  \citenamefont {Schmalian}, \citenamefont {Bud’ko}, \citenamefont
  {Canfield}, \citenamefont {Perakis},\ and\ \citenamefont
  {Wang}}]{Patz14_natcomm}%
  \BibitemOpen
  \bibfield  {author} {\bibinfo {author} {\bibfnamefont {A.}~\bibnamefont
  {Patz}}, \bibinfo {author} {\bibfnamefont {T.}~\bibnamefont {Li}}, \bibinfo
  {author} {\bibfnamefont {S.}~\bibnamefont {Ran}}, \bibinfo {author}
  {\bibfnamefont {R.~M.}\ \bibnamefont {Fernandes}}, \bibinfo {author}
  {\bibfnamefont {J.}~\bibnamefont {Schmalian}}, \bibinfo {author}
  {\bibfnamefont {S.~L.}\ \bibnamefont {Bud’ko}}, \bibinfo {author}
  {\bibfnamefont {P.~C.}\ \bibnamefont {Canfield}}, \bibinfo {author}
  {\bibfnamefont {I.~E.}\ \bibnamefont {Perakis}}, \ and\ \bibinfo {author}
  {\bibfnamefont {J.}~\bibnamefont {Wang}},\ }\href@noop {} {\bibfield
  {journal} {\bibinfo  {journal} {Nat Commun}\ }\textbf {\bibinfo {volume}
  {5}},\ \bibinfo {pages} {3229} (\bibinfo {year} {2014})}\BibitemShut
  {NoStop}%
\bibitem [{\citenamefont {Lv}\ \emph {et~al.}(2009)\citenamefont {Lv},
  \citenamefont {Wu},\ and\ \citenamefont {Phillips}}]{Lv09_prb}%
  \BibitemOpen
  \bibfield  {author} {\bibinfo {author} {\bibfnamefont {W.}~\bibnamefont
  {Lv}}, \bibinfo {author} {\bibfnamefont {J.}~\bibnamefont {Wu}}, \ and\
  \bibinfo {author} {\bibfnamefont {P.}~\bibnamefont {Phillips}},\ }\href@noop
  {} {\bibfield  {journal} {\bibinfo  {journal} {Phys. Rev. B}\ }\textbf
  {\bibinfo {volume} {80}},\ \bibinfo {pages} {224506} (\bibinfo {year}
  {2009})}\BibitemShut {NoStop}%
\bibitem [{\citenamefont {Lee}\ \emph {et~al.}(2009)\citenamefont {Lee},
  \citenamefont {Yin},\ and\ \citenamefont {Ku}}]{Lee09_prl}%
  \BibitemOpen
  \bibfield  {author} {\bibinfo {author} {\bibfnamefont {C.-C.}\ \bibnamefont
  {Lee}}, \bibinfo {author} {\bibfnamefont {W.-G.}\ \bibnamefont {Yin}}, \ and\
  \bibinfo {author} {\bibfnamefont {W.}~\bibnamefont {Ku}},\ }\href@noop {}
  {\bibfield  {journal} {\bibinfo  {journal} {Phys. Rev. Lett.}\ }\textbf
  {\bibinfo {volume} {103}},\ \bibinfo {pages} {267001} (\bibinfo {year}
  {2009})}\BibitemShut {NoStop}%
\bibitem [{\citenamefont {Chen}\ \emph {et~al.}(2010)\citenamefont {Chen},
  \citenamefont {Maciejko}, \citenamefont {Sorini}, \citenamefont {Moritz},
  \citenamefont {Singh},\ and\ \citenamefont {Devereaux}}]{Chen10_prb}%
  \BibitemOpen
  \bibfield  {author} {\bibinfo {author} {\bibfnamefont {C.-C.}\ \bibnamefont
  {Chen}}, \bibinfo {author} {\bibfnamefont {J.}~\bibnamefont {Maciejko}},
  \bibinfo {author} {\bibfnamefont {A.~P.}\ \bibnamefont {Sorini}}, \bibinfo
  {author} {\bibfnamefont {B.}~\bibnamefont {Moritz}}, \bibinfo {author}
  {\bibfnamefont {R.~R.~P.}\ \bibnamefont {Singh}}, \ and\ \bibinfo {author}
  {\bibfnamefont {T.~P.}\ \bibnamefont {Devereaux}},\ }\href@noop {} {\bibfield
   {journal} {\bibinfo  {journal} {Phys. Rev. B}\ }\textbf {\bibinfo {volume}
  {82}},\ \bibinfo {pages} {100504} (\bibinfo {year} {2010})}\BibitemShut
  {NoStop}%
\bibitem [{\citenamefont {Lee}\ and\ \citenamefont
  {Phillips}(2012)}]{Lee12_prb}%
  \BibitemOpen
  \bibfield  {author} {\bibinfo {author} {\bibfnamefont {W.-C.}\ \bibnamefont
  {Lee}}\ and\ \bibinfo {author} {\bibfnamefont {P.~W.}\ \bibnamefont
  {Phillips}},\ }\href@noop {} {\bibfield  {journal} {\bibinfo  {journal}
  {Phys. Rev. B}\ }\textbf {\bibinfo {volume} {86}},\ \bibinfo {pages} {245113}
  (\bibinfo {year} {2012})}\BibitemShut {NoStop}%
\bibitem [{\citenamefont {Liu}\ \emph {et~al.}(2011)\citenamefont {Liu},
  \citenamefont {Quan}, \citenamefont {Chen}, \citenamefont {Zou},\ and\
  \citenamefont {Lin}}]{ZouLJ_prb}%
  \BibitemOpen
  \bibfield  {author} {\bibinfo {author} {\bibfnamefont {D.-Y.}\ \bibnamefont
  {Liu}}, \bibinfo {author} {\bibfnamefont {Y.-M.}\ \bibnamefont {Quan}},
  \bibinfo {author} {\bibfnamefont {D.-M.}\ \bibnamefont {Chen}}, \bibinfo
  {author} {\bibfnamefont {L.-J.}\ \bibnamefont {Zou}}, \ and\ \bibinfo
  {author} {\bibfnamefont {H.-Q.}\ \bibnamefont {Lin}},\ }\href@noop {}
  {\bibfield  {journal} {\bibinfo  {journal} {Phys. Rev. B}\ }\textbf {\bibinfo
  {volume} {84}},\ \bibinfo {pages} {064435} (\bibinfo {year}
  {2011})}\BibitemShut {NoStop}%
\bibitem [{\citenamefont {Liang}\ \emph {et~al.}(2013)\citenamefont {Liang},
  \citenamefont {Moreo},\ and\ \citenamefont {Dagotto}}]{Liang13_prl}%
  \BibitemOpen
  \bibfield  {author} {\bibinfo {author} {\bibfnamefont {S.}~\bibnamefont
  {Liang}}, \bibinfo {author} {\bibfnamefont {A.}~\bibnamefont {Moreo}}, \ and\
  \bibinfo {author} {\bibfnamefont {E.}~\bibnamefont {Dagotto}},\ }\href@noop
  {} {\bibfield  {journal} {\bibinfo  {journal} {Phys. Rev. Lett.}\ }\textbf
  {\bibinfo {volume} {111}},\ \bibinfo {pages} {047004} (\bibinfo {year}
  {2013})}\BibitemShut {NoStop}%
\bibitem [{\citenamefont {Fang}\ \emph {et~al.}(2008)\citenamefont {Fang},
  \citenamefont {Yao}, \citenamefont {Tsai}, \citenamefont {Hu},\ and\
  \citenamefont {Kivelson}}]{Fang08_prb}%
  \BibitemOpen
  \bibfield  {author} {\bibinfo {author} {\bibfnamefont {C.}~\bibnamefont
  {Fang}}, \bibinfo {author} {\bibfnamefont {H.}~\bibnamefont {Yao}}, \bibinfo
  {author} {\bibfnamefont {W.-F.}\ \bibnamefont {Tsai}}, \bibinfo {author}
  {\bibfnamefont {J.}~\bibnamefont {Hu}}, \ and\ \bibinfo {author}
  {\bibfnamefont {S.~A.}\ \bibnamefont {Kivelson}},\ }\href@noop {} {\bibfield
  {journal} {\bibinfo  {journal} {Phys. Rev. B}\ }\textbf {\bibinfo {volume}
  {77}},\ \bibinfo {pages} {224509} (\bibinfo {year} {2008})}\BibitemShut
  {NoStop}%
\bibitem [{\citenamefont {Xu}\ \emph {et~al.}(2008)\citenamefont {Xu},
  \citenamefont {Müller},\ and\ \citenamefont {Sachdev}}]{Xu08_prb}%
  \BibitemOpen
  \bibfield  {author} {\bibinfo {author} {\bibfnamefont {C.}~\bibnamefont
  {Xu}}, \bibinfo {author} {\bibfnamefont {M.}~\bibnamefont {Müller}}, \ and\
  \bibinfo {author} {\bibfnamefont {S.}~\bibnamefont {Sachdev}},\ }\href@noop
  {} {\bibfield  {journal} {\bibinfo  {journal} {Phys. Rev. B}\ }\textbf
  {\bibinfo {volume} {78}},\ \bibinfo {pages} {020501} (\bibinfo {year}
  {2008})}\BibitemShut {NoStop}%
\bibitem [{\citenamefont {Fernandes}\ \emph {et~al.}(2014)\citenamefont
  {Fernandes}, \citenamefont {Chubukov},\ and\ \citenamefont
  {Schmalian}}]{Fernandes14_NatPhys}%
  \BibitemOpen
  \bibfield  {author} {\bibinfo {author} {\bibfnamefont {R.~M.}\ \bibnamefont
  {Fernandes}}, \bibinfo {author} {\bibfnamefont {A.~V.}\ \bibnamefont
  {Chubukov}}, \ and\ \bibinfo {author} {\bibfnamefont {J.}~\bibnamefont
  {Schmalian}},\ }\href@noop {} {\bibfield  {journal} {\bibinfo  {journal} {Nat
  Phys}\ }\textbf {\bibinfo {volume} {10}},\ \bibinfo {pages} {97} (\bibinfo
  {year} {2014})}\BibitemShut {NoStop}%
\bibitem [{\citenamefont {Fu}\ \emph {et~al.}(2012)\citenamefont {Fu},
  \citenamefont {Torchetti}, \citenamefont {Imai}, \citenamefont {Ning},
  \citenamefont {Yan},\ and\ \citenamefont {Sefat}}]{Fu12_prl}%
  \BibitemOpen
  \bibfield  {author} {\bibinfo {author} {\bibfnamefont {M.}~\bibnamefont
  {Fu}}, \bibinfo {author} {\bibfnamefont {D.~A.}\ \bibnamefont {Torchetti}},
  \bibinfo {author} {\bibfnamefont {T.}~\bibnamefont {Imai}}, \bibinfo {author}
  {\bibfnamefont {F.~L.}\ \bibnamefont {Ning}}, \bibinfo {author}
  {\bibfnamefont {J.-Q.}\ \bibnamefont {Yan}}, \ and\ \bibinfo {author}
  {\bibfnamefont {A.~S.}\ \bibnamefont {Sefat}},\ }\href@noop {} {\bibfield
  {journal} {\bibinfo  {journal} {Phys. Rev. Lett.}\ }\textbf {\bibinfo
  {volume} {109}},\ \bibinfo {pages} {247001} (\bibinfo {year}
  {2012})}\BibitemShut {NoStop}%
\bibitem [{\citenamefont {Zhang}\ \emph {et~al.}(2015)\citenamefont {Zhang},
  \citenamefont {Fernandes}, \citenamefont {Lamsal}, \citenamefont {Yan},
  \citenamefont {Chi}, \citenamefont {Tucker}, \citenamefont {Pratt},
  \citenamefont {Lynn}, \citenamefont {McCallum}, \citenamefont {Canfield},
  \citenamefont {Lograsso}, \citenamefont {Goldman}, \citenamefont {Vaknin},\
  and\ \citenamefont {McQueeney}}]{Zhang15_prl}%
  \BibitemOpen
  \bibfield  {author} {\bibinfo {author} {\bibfnamefont {Q.}~\bibnamefont
  {Zhang}}, \bibinfo {author} {\bibfnamefont {R.~M.}\ \bibnamefont
  {Fernandes}}, \bibinfo {author} {\bibfnamefont {J.}~\bibnamefont {Lamsal}},
  \bibinfo {author} {\bibfnamefont {J.}~\bibnamefont {Yan}}, \bibinfo {author}
  {\bibfnamefont {S.}~\bibnamefont {Chi}}, \bibinfo {author} {\bibfnamefont
  {G.~S.}\ \bibnamefont {Tucker}}, \bibinfo {author} {\bibfnamefont {D.~K.}\
  \bibnamefont {Pratt}}, \bibinfo {author} {\bibfnamefont {J.~W.}\ \bibnamefont
  {Lynn}}, \bibinfo {author} {\bibfnamefont {R.}~\bibnamefont {McCallum}},
  \bibinfo {author} {\bibfnamefont {P.~C.}\ \bibnamefont {Canfield}}, \bibinfo
  {author} {\bibfnamefont {T.~A.}\ \bibnamefont {Lograsso}}, \bibinfo {author}
  {\bibfnamefont {A.~I.}\ \bibnamefont {Goldman}}, \bibinfo {author}
  {\bibfnamefont {D.}~\bibnamefont {Vaknin}}, \ and\ \bibinfo {author}
  {\bibfnamefont {R.~J.}\ \bibnamefont {McQueeney}},\ }\href@noop {} {\bibfield
   {journal} {\bibinfo  {journal} {Phys. Rev. Lett.}\ }\textbf {\bibinfo
  {volume} {114}},\ \bibinfo {pages} {057001} (\bibinfo {year}
  {2015})}\BibitemShut {NoStop}%
\bibitem [{\citenamefont {Danilkin}\ \emph {et~al.}(2007)\citenamefont
  {Danilkin}, \citenamefont {Horton}, \citenamefont {Moore}, \citenamefont
  {Braoudakis},\ and\ \citenamefont {Hagen}}]{taipan}%
  \BibitemOpen
  \bibfield  {author} {\bibinfo {author} {\bibfnamefont {S.~A.}\ \bibnamefont
  {Danilkin}}, \bibinfo {author} {\bibfnamefont {G.}~\bibnamefont {Horton}},
  \bibinfo {author} {\bibfnamefont {R.}~\bibnamefont {Moore}}, \bibinfo
  {author} {\bibfnamefont {G.}~\bibnamefont {Braoudakis}}, \ and\ \bibinfo
  {author} {\bibfnamefont {M.}~\bibnamefont {Hagen}},\ }\href {\doibase
  10.1080/10238160601045755} {\bibfield  {journal} {\bibinfo  {journal}
  {Journal of Neutron Research}\ }\textbf {\bibinfo {volume} {15}},\ \bibinfo
  {pages} {55} (\bibinfo {year} {2007})}\BibitemShut {NoStop}%
\bibitem [{\citenamefont {Song}\ \emph {et~al.}(2013)\citenamefont {Song},
  \citenamefont {Regnault}, \citenamefont {Zhang}, \citenamefont {Tan},
  \citenamefont {Carr}, \citenamefont {Chi}, \citenamefont {Christianson},
  \citenamefont {Xiang},\ and\ \citenamefont {Dai}}]{Song13_prb}%
  \BibitemOpen
  \bibfield  {author} {\bibinfo {author} {\bibfnamefont {Y.}~\bibnamefont
  {Song}}, \bibinfo {author} {\bibfnamefont {L.-P.}\ \bibnamefont {Regnault}},
  \bibinfo {author} {\bibfnamefont {C.}~\bibnamefont {Zhang}}, \bibinfo
  {author} {\bibfnamefont {G.}~\bibnamefont {Tan}}, \bibinfo {author}
  {\bibfnamefont {S.~V.}\ \bibnamefont {Carr}}, \bibinfo {author}
  {\bibfnamefont {S.}~\bibnamefont {Chi}}, \bibinfo {author} {\bibfnamefont
  {A.~D.}\ \bibnamefont {Christianson}}, \bibinfo {author} {\bibfnamefont
  {T.}~\bibnamefont {Xiang}}, \ and\ \bibinfo {author} {\bibfnamefont
  {P.}~\bibnamefont {Dai}},\ }\href {\doibase 10.1103/PhysRevB.88.134512}
  {\bibfield  {journal} {\bibinfo  {journal} {Phys. Rev. B}\ }\textbf {\bibinfo
  {volume} {88}},\ \bibinfo {pages} {134512} (\bibinfo {year}
  {2013})}\BibitemShut {NoStop}%
\bibitem [{\citenamefont {Ma}\ \emph {et~al.}(2011)\citenamefont {Ma},
  \citenamefont {Chen}, \citenamefont {Yao}, \citenamefont {Zhang},
  \citenamefont {Zhang}, \citenamefont {Xia},\ and\ \citenamefont
  {Yu}}]{Ma11_prb}%
  \BibitemOpen
  \bibfield  {author} {\bibinfo {author} {\bibfnamefont {L.}~\bibnamefont
  {Ma}}, \bibinfo {author} {\bibfnamefont {G.~F.}\ \bibnamefont {Chen}},
  \bibinfo {author} {\bibfnamefont {D.-X.}\ \bibnamefont {Yao}}, \bibinfo
  {author} {\bibfnamefont {J.}~\bibnamefont {Zhang}}, \bibinfo {author}
  {\bibfnamefont {S.}~\bibnamefont {Zhang}}, \bibinfo {author} {\bibfnamefont
  {T.-L.}\ \bibnamefont {Xia}}, \ and\ \bibinfo {author} {\bibfnamefont
  {W.}~\bibnamefont {Yu}},\ }\href {\doibase 10.1103/PhysRevB.83.132501}
  {\bibfield  {journal} {\bibinfo  {journal} {Phys. Rev. B}\ }\textbf {\bibinfo
  {volume} {83}},\ \bibinfo {pages} {132501} (\bibinfo {year}
  {2011})}\BibitemShut {NoStop}%
\bibitem [{\citenamefont {Deng}\ \emph {et~al.}(2015)\citenamefont {Deng},
  \citenamefont {Liu}, \citenamefont {Xing}, \citenamefont {Yang},\ and\
  \citenamefont {Wen}}]{Deng15_prb}%
  \BibitemOpen
  \bibfield  {author} {\bibinfo {author} {\bibfnamefont {Q.}~\bibnamefont
  {Deng}}, \bibinfo {author} {\bibfnamefont {J.}~\bibnamefont {Liu}}, \bibinfo
  {author} {\bibfnamefont {J.}~\bibnamefont {Xing}}, \bibinfo {author}
  {\bibfnamefont {H.}~\bibnamefont {Yang}}, \ and\ \bibinfo {author}
  {\bibfnamefont {H.-H.}\ \bibnamefont {Wen}},\ }\href@noop {} {\bibfield
  {journal} {\bibinfo  {journal} {Phys. Rev. B}\ }\textbf {\bibinfo {volume}
  {91}},\ \bibinfo {pages} {020508} (\bibinfo {year} {2015})}\BibitemShut
  {NoStop}%
\bibitem [{\citenamefont {Rosenthal}\ \emph {et~al.}(2014)\citenamefont
  {Rosenthal}, \citenamefont {Andrade}, \citenamefont {Arguello}, \citenamefont
  {Fernandes}, \citenamefont {Xing}, \citenamefont {Wang}, \citenamefont {Jin},
  \citenamefont {Millis},\ and\ \citenamefont
  {Pasupathy}}]{Rosenthal14_natphys}%
  \BibitemOpen
  \bibfield  {author} {\bibinfo {author} {\bibfnamefont {E.~P.}\ \bibnamefont
  {Rosenthal}}, \bibinfo {author} {\bibfnamefont {E.~F.}\ \bibnamefont
  {Andrade}}, \bibinfo {author} {\bibfnamefont {C.~J.}\ \bibnamefont
  {Arguello}}, \bibinfo {author} {\bibfnamefont {R.~M.}\ \bibnamefont
  {Fernandes}}, \bibinfo {author} {\bibfnamefont {L.~Y.}\ \bibnamefont {Xing}},
  \bibinfo {author} {\bibfnamefont {X.~C.}\ \bibnamefont {Wang}}, \bibinfo
  {author} {\bibfnamefont {C.~Q.}\ \bibnamefont {Jin}}, \bibinfo {author}
  {\bibfnamefont {A.~J.}\ \bibnamefont {Millis}}, \ and\ \bibinfo {author}
  {\bibfnamefont {A.~N.}\ \bibnamefont {Pasupathy}},\ }\href
  {http://dx.doi.org/10.1038/nphys2870} {\bibfield  {journal} {\bibinfo
  {journal} {Nat. Phys.}\ }\textbf {\bibinfo {volume} {10}},\ \bibinfo {pages}
  {225} (\bibinfo {year} {2014})},\ \bibinfo {note} {article}\BibitemShut
  {NoStop}%
\bibitem [{\citenamefont {Cai}\ \emph {et~al.}(2014)\citenamefont {Cai},
  \citenamefont {Ruan}, \citenamefont {Zhou}, \citenamefont {Ye}, \citenamefont
  {Wang}, \citenamefont {Chen}, \citenamefont {Lee},\ and\ \citenamefont
  {Wang}}]{Cai14_prl}%
  \BibitemOpen
  \bibfield  {author} {\bibinfo {author} {\bibfnamefont {P.}~\bibnamefont
  {Cai}}, \bibinfo {author} {\bibfnamefont {W.}~\bibnamefont {Ruan}}, \bibinfo
  {author} {\bibfnamefont {X.}~\bibnamefont {Zhou}}, \bibinfo {author}
  {\bibfnamefont {C.}~\bibnamefont {Ye}}, \bibinfo {author} {\bibfnamefont
  {A.}~\bibnamefont {Wang}}, \bibinfo {author} {\bibfnamefont {X.}~\bibnamefont
  {Chen}}, \bibinfo {author} {\bibfnamefont {D.-H.}\ \bibnamefont {Lee}}, \
  and\ \bibinfo {author} {\bibfnamefont {Y.}~\bibnamefont {Wang}},\ }\href
  {\doibase 10.1103/PhysRevLett.112.127001} {\bibfield  {journal} {\bibinfo
  {journal} {Phys. Rev. Lett.}\ }\textbf {\bibinfo {volume} {112}},\ \bibinfo
  {pages} {127001} (\bibinfo {year} {2014})}\BibitemShut {NoStop}%
\bibitem [{\citenamefont {He}\ \emph {et~al.}(2010)\citenamefont {He},
  \citenamefont {Zhang}, \citenamefont {Xie}, \citenamefont {Wang},
  \citenamefont {Yang}, \citenamefont {Zhou}, \citenamefont {Chen},
  \citenamefont {Arita}, \citenamefont {Shimada}, \citenamefont {Namatame},
  \citenamefont {Taniguchi}, \citenamefont {Chen}, \citenamefont {Hu},\ and\
  \citenamefont {Feng}}]{He10_prl}%
  \BibitemOpen
  \bibfield  {author} {\bibinfo {author} {\bibfnamefont {C.}~\bibnamefont
  {He}}, \bibinfo {author} {\bibfnamefont {Y.}~\bibnamefont {Zhang}}, \bibinfo
  {author} {\bibfnamefont {B.~P.}\ \bibnamefont {Xie}}, \bibinfo {author}
  {\bibfnamefont {X.~F.}\ \bibnamefont {Wang}}, \bibinfo {author}
  {\bibfnamefont {L.~X.}\ \bibnamefont {Yang}}, \bibinfo {author}
  {\bibfnamefont {B.}~\bibnamefont {Zhou}}, \bibinfo {author} {\bibfnamefont
  {F.}~\bibnamefont {Chen}}, \bibinfo {author} {\bibfnamefont {M.}~\bibnamefont
  {Arita}}, \bibinfo {author} {\bibfnamefont {K.}~\bibnamefont {Shimada}},
  \bibinfo {author} {\bibfnamefont {H.}~\bibnamefont {Namatame}}, \bibinfo
  {author} {\bibfnamefont {M.}~\bibnamefont {Taniguchi}}, \bibinfo {author}
  {\bibfnamefont {X.~H.}\ \bibnamefont {Chen}}, \bibinfo {author}
  {\bibfnamefont {J.~P.}\ \bibnamefont {Hu}}, \ and\ \bibinfo {author}
  {\bibfnamefont {D.~L.}\ \bibnamefont {Feng}},\ }\href {\doibase
  10.1103/PhysRevLett.105.117002} {\bibfield  {journal} {\bibinfo  {journal}
  {Phys. Rev. Lett.}\ }\textbf {\bibinfo {volume} {105}},\ \bibinfo {pages}
  {117002} (\bibinfo {year} {2010})}\BibitemShut {NoStop}%
\bibitem [{\citenamefont {Zhang}\ \emph {et~al.}(2012)\citenamefont {Zhang},
  \citenamefont {He}, \citenamefont {Ye}, \citenamefont {Jiang}, \citenamefont
  {Chen}, \citenamefont {Xu}, \citenamefont {Ge}, \citenamefont {Xie},
  \citenamefont {Wei}, \citenamefont {Aeschlimann}, \citenamefont {Cui},
  \citenamefont {Shi}, \citenamefont {Hu},\ and\ \citenamefont
  {Feng}}]{Zhang12_prb}%
  \BibitemOpen
  \bibfield  {author} {\bibinfo {author} {\bibfnamefont {Y.}~\bibnamefont
  {Zhang}}, \bibinfo {author} {\bibfnamefont {C.}~\bibnamefont {He}}, \bibinfo
  {author} {\bibfnamefont {Z.~R.}\ \bibnamefont {Ye}}, \bibinfo {author}
  {\bibfnamefont {J.}~\bibnamefont {Jiang}}, \bibinfo {author} {\bibfnamefont
  {F.}~\bibnamefont {Chen}}, \bibinfo {author} {\bibfnamefont {M.}~\bibnamefont
  {Xu}}, \bibinfo {author} {\bibfnamefont {Q.~Q.}\ \bibnamefont {Ge}}, \bibinfo
  {author} {\bibfnamefont {B.~P.}\ \bibnamefont {Xie}}, \bibinfo {author}
  {\bibfnamefont {J.}~\bibnamefont {Wei}}, \bibinfo {author} {\bibfnamefont
  {M.}~\bibnamefont {Aeschlimann}}, \bibinfo {author} {\bibfnamefont {X.~Y.}\
  \bibnamefont {Cui}}, \bibinfo {author} {\bibfnamefont {M.}~\bibnamefont
  {Shi}}, \bibinfo {author} {\bibfnamefont {J.~P.}\ \bibnamefont {Hu}}, \ and\
  \bibinfo {author} {\bibfnamefont {D.~L.}\ \bibnamefont {Feng}},\ }\href
  {\doibase 10.1103/PhysRevB.85.085121} {\bibfield  {journal} {\bibinfo
  {journal} {Phys. Rev. B}\ }\textbf {\bibinfo {volume} {85}},\ \bibinfo
  {pages} {085121} (\bibinfo {year} {2012})}\BibitemShut {NoStop}%
\bibitem [{\citenamefont {Yi}\ \emph {et~al.}(2012)\citenamefont {Yi},
  \citenamefont {Lu}, \citenamefont {Moore}, \citenamefont {Kihou},
  \citenamefont {Lee}, \citenamefont {Iyo}, \citenamefont {Eisaki},
  \citenamefont {Yoshida}, \citenamefont {Fujimori},\ and\ \citenamefont
  {Shen}}]{Yi12_njp}%
  \BibitemOpen
  \bibfield  {author} {\bibinfo {author} {\bibfnamefont {M.}~\bibnamefont
  {Yi}}, \bibinfo {author} {\bibfnamefont {D.~H.}\ \bibnamefont {Lu}}, \bibinfo
  {author} {\bibfnamefont {R.~G.}\ \bibnamefont {Moore}}, \bibinfo {author}
  {\bibfnamefont {K.}~\bibnamefont {Kihou}}, \bibinfo {author} {\bibfnamefont
  {C.-H.}\ \bibnamefont {Lee}}, \bibinfo {author} {\bibfnamefont
  {A.}~\bibnamefont {Iyo}}, \bibinfo {author} {\bibfnamefont {H.}~\bibnamefont
  {Eisaki}}, \bibinfo {author} {\bibfnamefont {T.}~\bibnamefont {Yoshida}},
  \bibinfo {author} {\bibfnamefont {A.}~\bibnamefont {Fujimori}}, \ and\
  \bibinfo {author} {\bibfnamefont {Z.-X.}\ \bibnamefont {Shen}},\ }\href
  {http://stacks.iop.org/1367-2630/14/i=7/a=073019} {\bibfield  {journal}
  {\bibinfo  {journal} {New J. Phys.}\ }\textbf {\bibinfo {volume} {14}},\
  \bibinfo {pages} {073019} (\bibinfo {year} {2012})}\BibitemShut {NoStop}%
\end{thebibliography}


\end{document}